\documentclass[twocolumn,secnumarabic,amssymb, nobibnotes, aps, pra,nofootinbib]{revtex4-2}

\pdfoutput=1

\newcommand{\env}[1]{\texttt{#1}}
\usepackage[colorlinks=true,citecolor=blue,allcolors=blue]{hyperref}  
\usepackage{graphicx}
\usepackage{siunitx}
\usepackage{amsmath}
\usepackage{tabularx}
\DeclareMathOperator\erfc{erfc}
\usepackage{hyperref}

\begin{abstract}
Established techniques for deterministically creating dark solitons in repulsively interacting atomic Bose-Einstein condensates (BECs) can only access a narrow range of soliton velocities. Because velocity affects the stability of individual solitons and the properties of soliton-soliton interactions, this technical limitation has hindered experimental progress.
Here we create dark solitons in highly anisotropic cigar-shaped BECs with arbitrary position and velocity by simultaneously engineering the amplitude and phase of the condensate wavefunction, improving upon previous techniques which explicitly manipulated only the condensate phase.
The single dark soliton solution present in true one-dimensional (1D) systems corresponds to the kink soliton in anisotropic three-dimensional systems and is joined by a host of additional dark solitons, including vortex ring and solitonic vortex solutions. 
We readily create dark solitons with speeds from zero to half the sound speed. The observed soliton oscillation frequency suggests that we imprinted solitonic vortices, which for our cigar-shaped system are the only stable solitons expected for these velocities. Our numerical simulations of 1D BECs show this technique to be equally effective for creating kink solitons when they are stable.
We demonstrate the utility of this technique by deterministically colliding dark solitons with domain walls in two-component spinor BECs.
\end{abstract}

\begin{document}
\title{Creating solitons with controllable and near-zero velocity in Bose-Einstein condensates}%

\author{A.~R.~Fritsch}
\author{Mingwu~Lu}
\author{G.~H.~Reid}
\author{A.~M.~Pi\~neiro}
\author{I.~B.~Spielman}
\email{ian.spielman@nist.gov}
\homepage{http://ultracold.jqi.umd.edu}
\affiliation{Joint Quantum Institute, National Institute of Standards and Technology, and University of Maryland, Gaithersburg, Maryland, 20899, USA}
\maketitle

Localized excitations such as vortices and kink solitons in atomic Bose-Einstein condensates (BECs) have attracted much attention. Solitons are long-lived shape-preserving solitary waves typically stabilized by a balance between linear and non linear effects \cite{PhysRevLett.89.110401,PhysRevA.70.043604,PhysRevLett.118.190401,Juzeli_nas_2007,PhysRevA.96.013612}.
Experiments with dark solitons in repulsively interacting systems first identified different regimes of stability for solitons~\cite{PhysRevLett.83.5198,Denschlag97,PhysRevLett.86.2926} before exploring effects including soliton interactions~\cite{Becker2008} and diffusion~ \cite{Aycock2503}.  For all their success, these experiments could deterministically produce solitons within only a narrow  range of velocities.  Inspired by Ref. \cite{PhysRevA.63.051601}, we demonstrate a straightforward technique for creating dark solitons that provides full control over their position and velocity.

The propagation velocity of dark solitons in quasi-one-dimensional (1D) systems determines many of their properties; indeed, their maximum speed is bounded by the local speed of sound $c$ in their host BEC. A pair of colliding dark solitons will, long after the collision, have unchanged shape and velocity, with the effect of interaction being only in a displacement in their asymptotic trajectories~\cite{Frantzeskakis_2010}. For sufficiently small relative velocities the collision can be viewed as if the two solitons reflect from each other like equal-mass hard spheres~\cite{PhysRevLett.101.130401}, giving the same asymptotic velocity and shapes as if they had passed through each other. This is sometimes referred to as a ``non interacting" collision.

A dark soliton manifests as a moving density depletion in a BEC (we call a stationary density depletion a ``black soliton"), whose width increases with velocity and whose depth decreases with velocity. Across a soliton, the underlying BEC wavefunction has a phase difference that determines its velocity. In anisotropic 3D systems such as ours, this overall longitudinal phase drop can be accompanied by transverse structure, leading to a host of distinct dark solitons.  The single dark soliton solution present in true 1D systems, for which a transverse structure is absent, corresponds to the kink soliton in anisotropic three-dimensional (3D) systems.  The solitonic vortex is another example of a solitonic excitation present in 3D systems; although its velocity-dependent longitudinal phase drop and density profile are qualitatively similar to those of kink solitons, as a vortex, it also has a phase-singularity where the density vanishes. In three-dimensional traps, kink solitons are stable only for a range of velocities that depend on the trap geometry~\cite{PhysRevLett.89.110401}. In regimes where kink solitons are unstable, solitonic vortices are stable.  For our system, the kink soliton is predicted to be unstable except for velocities very close to $c$, and for almost all parameters, the solitonic vortex is predicted to be the only stable solitonic excitation~\cite{Mateo_2015}.  

There is no single and universally agreed upon definition of a solitonic vortex in the literature. One definition hinges on the fact that a vortex in a channel whose transverse size is much smaller than its length and has the same asymptotic phase profile as a soliton in the longitudinal direction \cite{PhysRevLett.113.065301}. While the other definition notes that when the transverse harmonic oscillator length becomes comparable to the healing length, the vortex density profile is reminiscent of a kink soliton~\cite{Brand_2001}. In this paper we remain agnostic about the choice of definition, but our experiment corresponds to the first definition.

The established technique for producing dark solitons simply consists of laser-imprinting a longitudinal phase difference onto the BEC wavefunction. A soliton at rest typically has a density profile width of about $\SI{0.5}{\mu m}$, while its phase profile is a step function. Imaging systems used to optically imprint the phase change typically have resolution limits of $\SI{1.5}{\mu m}$ or more. Furthermore, the established technique does not reduce the density in a controllable way at the soliton position. Instead, the optical dipole force resulting from intensity change at the soliton position changes the density in a way that is not generally consistent with the imprinted phase difference. All these features give poor overlap with the desired soliton wave function~\cite{PhysRevLett.83.5198,Denschlag97,Aycock2503}. As a result, these experiments have only successfully produced solitons moving at a  substantial fraction of the sound velocity. We overcome this limit by simultaneously engineering the BEC density and phase.  This control allows us to improve the spatial mode-matching between the applied potential and the desired soliton wave function. This allows us to create solitons with speed ranging from zero to half the velocity of sound.

This manuscript is organized as follows: First, we summarize the essential properties of dark solitons in repulsively interacting 1D BECs, which still captures the essential physics for dark solitons in anisotropic 3D systems; second, we describe our experimental methods for imprinting and detecting solitons; third, we demonstrate our protocol and explore its range of applicability; and lastly, we exhibit the utility of this approach by colliding a dark soliton with a domain wall in a two-component spinor BEC.
\section{Dark Solitons in repulsive BECs}
Bose-Einstein condensates well below their transition temperatures, i.e., with negligible thermal fraction, are well described by the Gross-Pitaevskii equation (GPE). Solitons are exact solutions of the one-dimensional GPE~\cite{KIVSHAR2001225,Barenghi_2016},
\begin{equation}
i\hbar \frac{\partial}{\partial t}\Psi(z,t) = \left[-\frac{\hbar^2}{2m}\frac{\partial ^2 }{\partial z^2}+U(z)+g_{\rm 1D}|\Psi(z,t)|^2\right] \Psi(z,t),
\end{equation}
for particles of mass $m$, which approximately describes highly anisotropic systems where transverse motion is frozen out by strong confinement, leaving behind only longitudinal dynamics. For harmonic transverse confinement with transverse trapping frequencies $\omega_x$ and $\omega_y$, the 1D interaction constant is $g_{\rm 1D} = 2\hbar a (\omega_x\omega_y)^{1/2} $, where $a$ is the 3D $s$-wave scattering length. In the GPE, $\Psi(z,t)$ is interpreted as the condensate wave function whose magnitude gives the local 1D atomic density $n(z,t) = \vert \Psi(z,t)\vert ^2$. An infinite homogeneous system is characterized by the chemical potential $\mu =g_{\rm 1D}n$ and has low-energy phonon excitations with speed of sound $c = \sqrt{g_{\rm 1D}n / m}$~\cite{PhysRevA.57.518}.

A dark soliton appears as a moving depletion in the background BEC; for an infinite and homogeneous system a dark soliton with velocity $v$ is exactly described by
\begin{align}
\Psi_{\rm s}(z,t) &= \sqrt{n} \left\{i\frac{v}{c}+\sqrt{1-\frac{v^2}{c^2}}\right. \nonumber \\ 
& \left. \times\tanh \left[\sqrt{1-\frac{v^2}{c^2}}\frac{(z-v t)}{\xi}\right]\right\} \exp\left(-\frac{i\mu t}{\hbar}\right).
\end{align}
The healing length $\xi = \hbar/m c$ is the typical minimum length scale on which the condensate wavefunction can appreciably change. 
This soliton solution can be characterized by $\phi$, the change in the condensate phase across the soliton. The soliton velocity, characteristic density profile width $w_{\rm s}$, and depth $n_{\rm s}$ are all derived from the phase via
\begin{subequations}
    \begin{align}
      \frac{v}{c} &= \cos\frac{\phi}{2}, \\
      \frac{w_{\rm s}}{\xi} &= \csc\frac{\phi}{2} = \frac{1}{\sqrt{1-v^2/c^2}},  \\
      \frac{n_{\rm s}}{n} &= \sin^2 \frac{\phi}{2} = 1-\frac{v^2}{c^2}.
    \end{align}
  \end{subequations}
The phase change across a stationary black soliton, i.e., with zero central density, is therefore $\pi$. This result further shows that as the soliton velocity approaches that of sound, its wavefunction smoothly connects to the ground-state wave function.

Like most experiments with solitons, ours takes place in a highly elongated system with $\omega_z\ll\omega_{x,y}$, but with $\omega_{x,y} < \mu/\hbar$, requiring the 3D GPE for a proper description.  The 3D GPE supports many solitonic solutions~\cite{Mateo_2015}, from kink solitons---the analog to dark solitons in 1D---to vortex rings and solitonic vortices. Reference~\cite{PhysRevLett.89.110401} showed that with sufficient transverse confinement, kink solitons can be unconditionally stable in anisotropic 3D systems; however, our system is in a regime where only rapidly moving kink solitons with $v/c \gtrsim 0.9$ are stable.

The most common method to create solitons writes the phase drop $\phi$ associated with the soliton wave function onto initially homogeneous BECs. In practice, this phase is imprinted by briefly applying an external potential $V(z)$, for example, a step function, that changes the wave function phase by $\phi = V(z)t_{\rm p}/\hbar$, where $t_{\rm p}$ is the duration of the applied pulse. Typically, $V(z)$ is generated by a far-detuned laser, and $t_{\rm p}$ should be shorter than $t_{\rm c} = \hbar/\mu$, the time it takes an excitation moving at the speed of sound to traverse a single healing length, so the density remains unchanged over the pulse duration. 
Ideally, the imprinted phase would determine the initial soliton velocity; however, the phase imprinting system always has a finite resolution limiting the lowest possible soliton velocity. In our experiment $c\approx \SI{2}{mm/s}$ giving $\xi \approx \SI{0.4}{\mu m}$, so that a slowly moving soliton with velocity $0.1c$ will have a width $w_{\rm s}\approx \SI{0.4}{\mu m}$, which is much smaller than our resolution of $\SI{2.8}{\mu m}$.\footnote{\label{note1}Defined as the distance between the central peak and the first minimum of the Airy disk.} This rightly suggests that a BEC following such a phase imprinting process will differ significantly from the desired soliton wave function.

Our improved method overcomes this limitation by first depleting the density at the  desired soliton location with two important outcomes: (1) the imprinted phase profile can better match that of a soliton because decreasing the density locally increases the healing length, thereby increasing the soliton width $w_{\rm s}$; and (2) the density depletion is better mode matched to the density depletion at a soliton's center.
\section{Experimental system}
\begin{figure}[]
	\centering
	\includegraphics{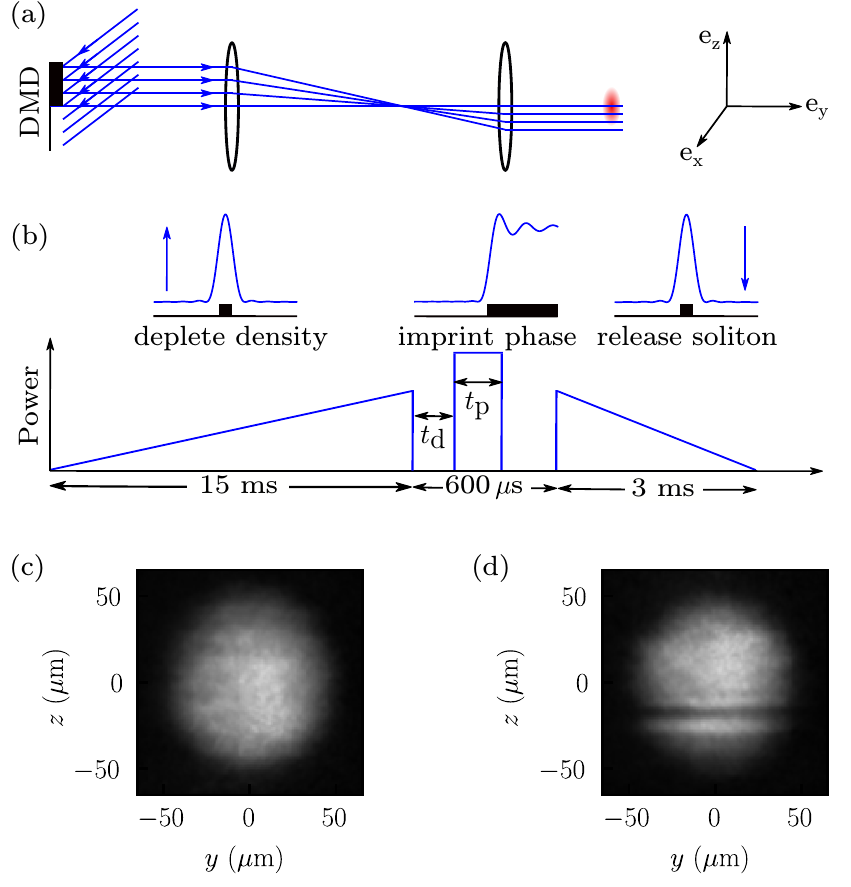}
	\caption{Experimental concept. (a) Far-detuned laser light illuminates the surface of a DMD, which is  programmed to reflect the light with the desired pattern. The DMD patterns are demagnified and imaged onto the atoms, which are represented as the red cloud. The trap laser beams are not shown in this figure. (b) Time sequence used to create solitons with the DMD pattern. The laser power during each phase of the experiment (vertical axis) was controlled using an acousto-optic modulator. The insets depict the potential resulting from the DMD patterns, accounting for the $\SI{2.8}{\mu m}$ (see footnote \ref{note1}) resolution of our imaging system (because of the finite aperture, but neglecting any aberrations). (c) Absorption image of a typical BEC after TOF without solitons. (d) A BEC created with similar experimental conditions including a soliton.}
\label{fig:fig1}
	\end{figure}

Our experiments begin with $^{87}\text{Rb}$ BECs having $N =\SI{2.4(2)\times 10^5}{atoms}$\footnote{All uncertainties herein reflect the uncorrelated combination of single-sigma statistical and systematic uncertainties.} in the $\vert f = 1, m_f = 0\rangle$ internal state in a time-averaged crossed optical dipole trap of wavelength $\lambda = \SI{1064}{n m}$. One of the beams is rastered along ${\bf e}_z$, with maximum displacement of about one waist, painting an elongated trap with frequencies $(\omega_x,\omega_y,\omega_z) = 2\pi\times\SI{[94.5(6),153(1),9.1(1)]}{Hz}$. The measured (see below) longitudinal Thomas-Fermi (TF) radius of our BEC is $R_{\text{TF}}^{z} = \SI{55(1)}{\mu m}$, and the chemical potential is $\mu = h\times\SI{1.1(1)}{\kilo \Hz}$. We create arbitrary repulsive potentials $V(z)$ using far blue-detuned laser light of wavelength $\lambda = \SI{777.6}{\nm}$, spatially patterned by a digital micromirror device (DMD, Texas Instruments DLP\textsuperscript{\textregistered} LightCrafter\textsuperscript{\texttrademark} Module, DLP3000).\footnote{Certain commercial equipment, instruments, or materials are identified in this paper in order to specify the experimental procedure adequately. Such identification is not intended to imply recommendation or endorsement by the National Institute of Standards and Technology, nor is it intended to imply that the materials or equipment identified are necessarily the best available for the purpose.} The light reflected by the DMD is imaged onto the atoms (the DMD surface is focused at the BEC) using an imaging system that demagnifies ($\times$12) the pattern as schematically shown in Fig. \ref{fig:fig1}\textcolor{blue}{(a)}. There we illustrate the situation where the DMD is programmed to reflect half of the laser profile.
We engineer the local BEC density with a dimple like potential created by programming the DMD to reflect the light in a stripe which, after demagnification, is about $\SI{270}{\mu m}$ wide along ${\bf e}_y$ and $\SI{1.8}{\mu m}$ along ${\bf e}_z$, as shown in the first inset to Fig. \ref{fig:fig1}\textcolor{blue}{(b)}, including the impact of finite resolution. We ramp the optical potential from zero to $h\times\SI{0.78(4)}{\kilo \Hz}$ in $\SI{15}{ms}$, reducing the local atomic density by about 70 $\%$. Next, the light is extinguished for $t_{\rm d} = \SI{100}{\mu s}$, while the DMD is updated\footnote{This {DMD} can refresh every $\SI{250}{\mu s}$, which includes about $\SI{150}{\mu s}$ of signal and/or electronics delay and about $\SI{100}{\mu s}$ for the mechanical response of the mirrors. Therefore, we command the DMD to change $\SI{150}{\mu s}$ before extinguishing the dimple light, and we wait another $\SI{100}{\mu s}$ while the mirrors change mechanically before applying the phase imprinting light. We leave the phase imprinting light on for the desired time. We anticipate reapplying the dimple by commanding the DMD to change to the dimple configuration at an appropriate time sufficiently long before reapplying the dimple light. The light is turned on/off in much less than $\SI{1}{\mu s}$ using an acousto-optic modulator.} to display a step function, illuminating the BEC to one side of the dimple. The step potential is applied for a time $t_{\rm p}$ up to $\SI{170}{\mu s}$, which, for our potential of magnitude V$ = h\times\SI{5.5(3)}{\kilo \Hz}$, results in an accumulated relative phase up to $\SI{1.9(1)}{\pi}$. To avoid abrupt changes in the BEC density after phase imprinting, we reapply the dimple potential and ramp its magnitude to zero in $\SI{3}{ms}$. The times to ramp the dimple up and down were chosen to be slow enough to be reasonably adiabatic and fast enough so that the soliton does not propagate very far during the ramp down. For comparison we create solitons without density engineering by applying the same time sequence with the dimple potential set to zero.

After a variable evolution time, the dipole trap potential is removed, allowing the BEC to expand for a $\SI{15}{ms}$ time of flight (TOF), after which time the BEC is imaged using standard absorption imaging \cite{ketterle1999making}. A typical image of a TOF-expanded BEC with no soliton imprinted is shown in Fig. \ref{fig:fig1}\textcolor{blue}{(c)}; each dark soliton appears as a local dip in the density as in Fig. \ref{fig:fig1}\textcolor{blue}{(d)}. Note that because of the trap geometry, the cloud in the ${\bf e}_y$ direction expands faster than in ${\bf e}_z$, such that the two directions have similar sizes at this particular expansion time. Because the BEC expands so slowly in the ${\bf e}_z$ direction, our analysis of the TOF images assumes that the $z$ positions in TOF correspond to the $z$ positions in trap at the time of release.

We experimentally calibrate the phase imprinting and density engineering potential by adiabatically loading the BEC into the step like potential usually used for phase imprinting described above. From the time-of-flight images and the Castin-Dum scaling theory of BEC expansion~\cite{PhysRevLett.77.5315} we determine the difference in the mean-field energy, i.e., the local chemical potential $\mu(z)$ between the two sides of the step like potential. This calibrates the step potential. For more details and discussion about the approximation involved, see Appendix \ref{app:pot}.

We measure the sound velocity in the BEC by intentionally inducing density perturbations in the condensate launched by abruptly turning on the dimple potential at the trap center, creating a pair of density waves traveling in opposite directions at the speed of sound~\cite{PhysRevLett.79.553}. We associate the average speed of these waves with a 1D speed of sound, giving $c = \SI{1.65(5)}{mm/s}$, which is related to the 3D speed of sound by $ c = c_{\rm 3D}/ \sqrt{2}$, as is appropriate for highly elongated 3D BECs such as ours \cite{PhysRevA.57.518}. This speed is consistent with the 3D speed of sound derived from the chemical potential $c_{\rm 3D} = \SI{2.2(1)}{mm/s}$, which would imply $c = \SI{1.58(7)}{mm/s} $.

Even our improved protocol does not result in the perfect soliton wave function, resulting in undesired excitations that manifest as density modulations propagating at the speed of sound. These reach the extremes of the BEC and dissipate in less than $\SI{100}{ms}$. We therefore take our earliest data $\SI{150}{ms}$ after phase imprinting to obtain a clean background for tracking solitons.   
\section{Results}
\begin{figure}[]
	\centering
	\includegraphics{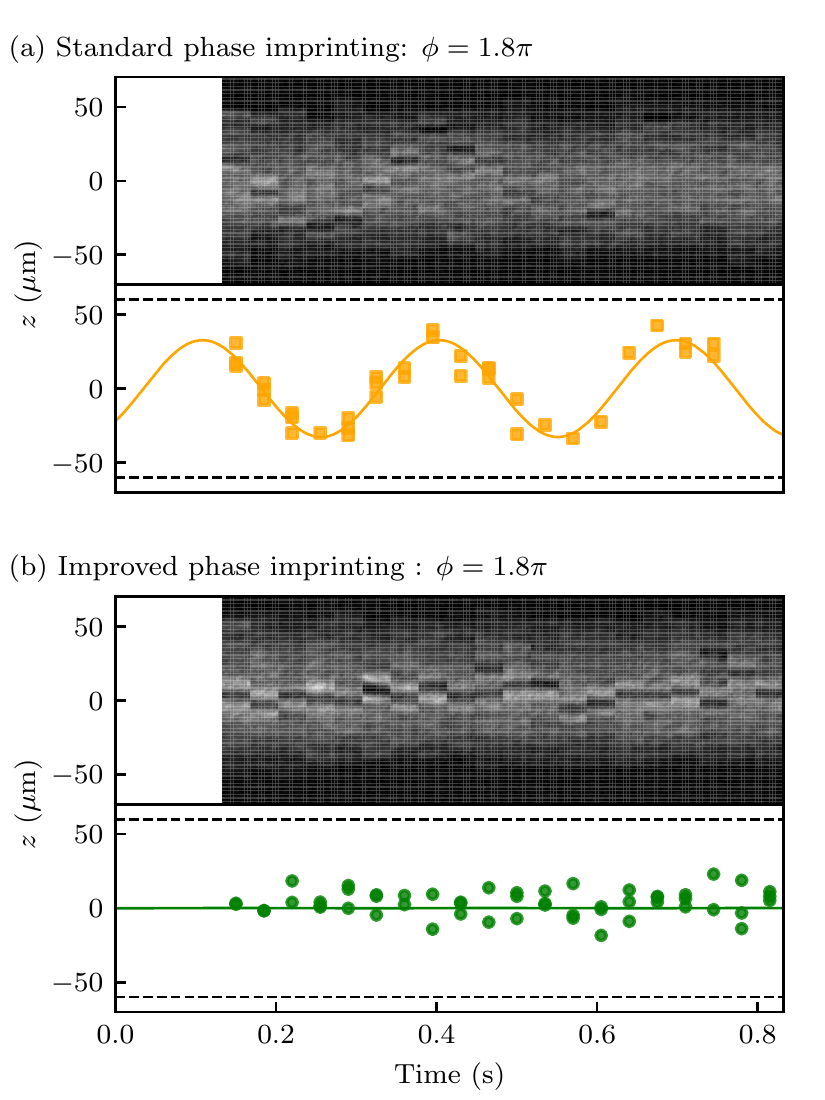}
	\caption{Oscillation of dark solitons created using a $\SI{1.8(1)}{\pi}$ phase imprint showing: (a) the standard protocol and (b) the improved protocol. In both cases, the top panel plots 1D longitudinal slices taken through our TOF-expanded BECs as a function of soliton evolution time for a single realization of the experiment at each time.  The bottom panels plot the resulting soliton positions (symbols), with three realizations of the experiment for each time.  The curve in (a) is a sinusoidal fit to the data described in the main text. Because the data in (b) reveals a stationary soliton, we used the procedure described in Appendix \ref{app:fft} to plot a sinusoidal oscillation with amplitude $A = \SI{0.19}{\mu m}$ and $\omega_{\rm s} =  2\pi\times \SI{3.0}{Hz}$ . Dashed lines represent the edges of our BECs.}
\label{fig:fig2}
	\end{figure}

Here we compare the standard phase imprinting protocol with our improved protocol for creating solitons.  Figure \ref{fig:fig2} illustrates our main result: (a) the standard protocol can only create solitons with a velocity so large that they have an oscillation amplitude comparable to the Thomas-Fermi radius; (b) the improved protocol can create solitons with no discernible motion, i.e, black solitons. In this section, we quantitatively compare the ability of these two techniques to create solitons on demand. 

We first turn to a more detailed discussion of the data presented in Fig. \ref{fig:fig2}, for which the phase imprinted was $\phi = \SI{1.8(1)}{\pi}$.  Both (a) and (b) include 1D cross-sectional slices of our BEC after TOF as a function of time, illustrating the representative density profiles from which we obtain the soliton position by fitting the dip to a Gaussian function. These positions, for three repetitions of the experiment, are displayed by the semitransparent symbols in the accompanying panel and follow approximately sinusoidal trajectories. This raw data shows key differences between solitons created via these two protocols: the shallower solitons created using the standard protocol oscillate with larger amplitude, implying a higher peak velocity. This is consistent with our expectations that more rapidly moving solitons are associated with a shallower density depletion.

In Fig. \ref{fig:fig2}, when fewer than three symbols are displayed, no soliton was observed in one or more of the trials at that time. Our data therefore indicates that fast solitons, created using the standard protocol, have shorter lifetimes than the stationary solitons. We suspect that this results in part from friction dissipation mechanisms, leading to a more rapid destabilization of fast-moving solitons similar to those discussed in Refs.~\cite{PhysRevLett.116.225301,Aycock2503,Hurst2017}. In the latter two references, diffusion and damping resulted from the interaction with a dilute background of impurities.  In the present case, the computation of the collision integral would result from the reflection of phonons rather than impurities.

For harmonically trapped 1D BECs, solitons follow sinusoidal trajectories described by $z(t) = A\sin(\omega_{\rm s} t + \phi)$, with predicted oscillation frequency $\omega_{\rm sol} = \omega_z/\sqrt{2}$, which is also valid for kink solitons in 3D. The factor of $\sqrt{2}$ is often described in terms of the ratio between the ``inertial" and ``physical" (or ``bare") masses of a dark soliton, $\eta = m_{\rm in}/m_{\rm ph}$,  $\omega_{\rm sol} = \omega_z/\sqrt{\eta}$, with $\eta_{\rm k} = 2$ for a kink soliton. The bare mass is related to the number of missing particles, and the inertial mass is related to the response to an external force. We measure an oscillation frequency of $\omega_{\rm s} =  2\pi\times \SI{3.0(1)}{Hz} $. This differs significantly from the expected $\omega_z/\sqrt{2}= 2\pi\times6.4(1)\ {\rm Hz}$ for a kink soliton. Therefore, we consider the possibility that we are observing a solitonic vortex. The mass ratio for solitonic vortices exhibiting small-amplitude oscillations depends on the chemical potential and transverse trapping frequency $\omega_{\perp}$~\cite{Mateo_2015}, which for our case gives $\eta_{\rm sv} \approx 9$. For $\eta_{\rm sv} \approx 9$ the computed oscillation frequency is $\omega_{\rm s} \approx  2\pi\times \SI{3}{Hz}$, consistent with the observed frequency. Because our transverse frequencies are unequal, it is not \textit{a priori} clear how to average them, or if averaging is suitable at all.  Table \ref{tab:tab1} displays the transverse trapping frequencies and possible averages; we find both the geometric and arithmetic averages to be consistent with the observed soliton oscillation frequency.
\begin{table}[]
\centering
\begin{tabular*}{\columnwidth}{@{\extracolsep{\fill}}clc}
\toprule
\multicolumn{2}{c}{${\omega_{\perp}}$}                       &                                            \\
Expression                    & \multicolumn{1}{c}{Value} & \multicolumn{1}{l}{$\eta_{\rm sv}\approx$} \\ \hline
$\omega_x$                    & $2\pi\times 94.5(6)$ Hz   & $13$                                       \\
$\sqrt{\omega_x  \omega_y}$   & $2\pi\times 120.2(5)$ Hz  & $ 9$                                       \\
$(\omega_x  + \omega_y) / 2 $ & $2\pi\times 123.8(6)$ Hz  & $ 9$                                       \\
$\omega_y $                   & $2\pi\times 153(1)$ Hz  & $7$                                        \\ 
\toprule
\end{tabular*}
\caption{Transverse trapping frequencies $\omega_{\perp}$ and typical averages and the associated solitonic vortex mass ratio.}
\label{tab:tab1}
\end{table}

\begin{figure}[]
	\centering
	\includegraphics{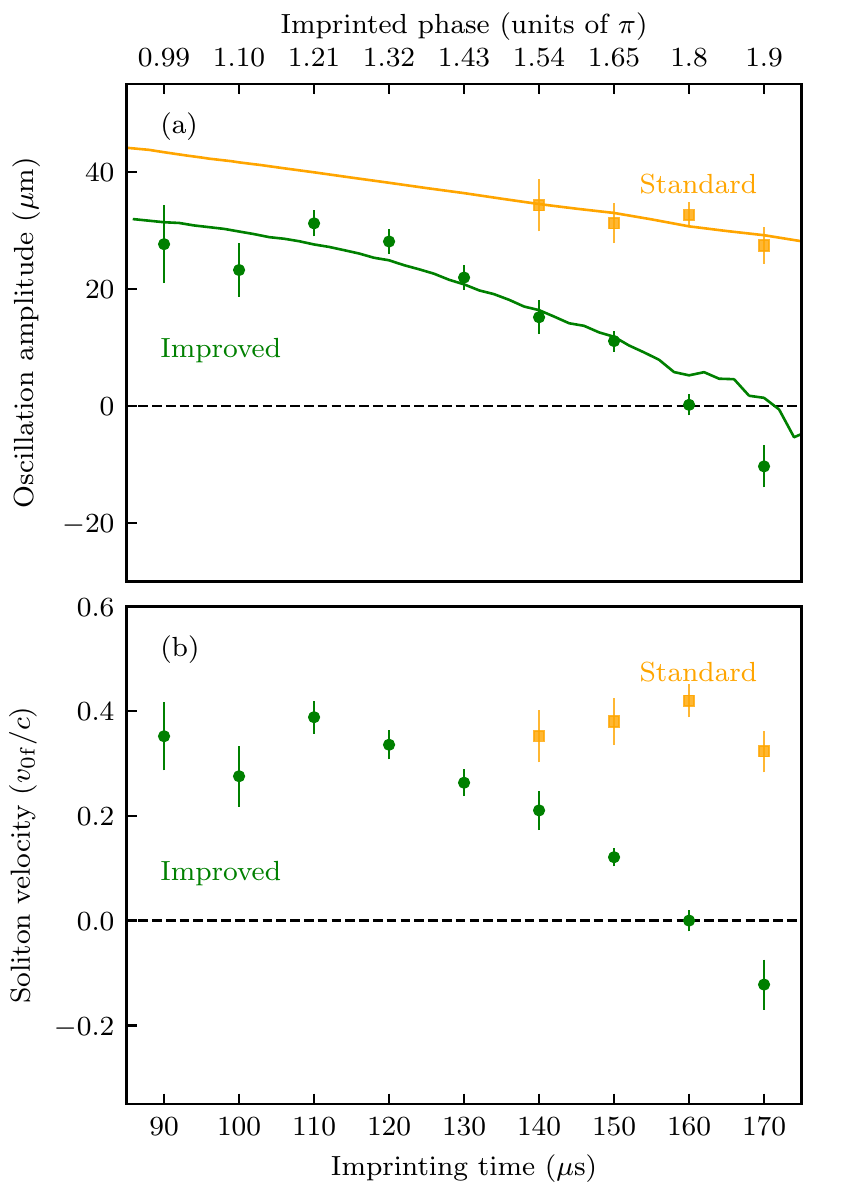}
	\caption{Soliton oscillation amplitude (a) and velocity (b) for different imprinted phases. The imprinted phase (top axis) is related to the imprinting time (bottom axis) by our calibration of the DMD potential. Solid curves show the results of numerical simulations, and filled symbols are experimental data for the standard (orange) and improved (green) protocols. Error bars represent the standard deviation from the sinusoidal fits.}
\label{fig:fig3}
	\end{figure}

Figure \ref{fig:fig3}\textcolor{blue}{(a)} plots the TOF soliton oscillation amplitude, obtained from a sinusoidal fit (except for imprinting time $\SI{160}{\mu s}$, where we used the procedure described in Appendix \ref{app:fft}), as a function of the imprinting phase, and compares our experimental data (filled symbols) to results of 1D GPE simulations (solid curves, see Appendix \ref{app:gpe}), which we expect to provide only qualitative guidance for solitonic vortices. This figure shows that the standard protocol (orange) creates solitons within a very narrow window of amplitudes, all comparable to $R_{\rm TF}$, while the improved protocol (green) can tune the oscillation amplitude from this large value through zero. We observe experimentally that visible solitons are not created for phase imprints below about $\pi$, while above $2\pi$, multiple solitons are created. Figure \ref{fig:fig3} therefore focuses on phase imprinting within this interval. The velocities, extracted from the sinusoidal fits and normalized by our measured sound velocity, are plotted in Fig. \ref{fig:fig3}\textcolor{blue}{(b)}. This data shows that when standard phase imprinting is used (orange squares), solitons can only be created with velocities around ${0.4c}$, which is consistent with previous works \cite{Becker2008,PhysRevLett.101.120406}. In contrast, when the improved method is used (green circles) we acquire almost full control over the soliton velocity, including the ability to generate nearly stationary solitons. In this case, the oscillation amplitude is $A_{\rm I} = \SI{0(2)}{\mu m}$, resulting in a peak velocity of ${0.00(4)c}$, which corresponds to a stationary soliton within our experimental uncertainty (see Appendix \ref{app:fft}).
\section{Soliton-domain wall collisions}
We conclude with an application of our technique, deterministically colliding a dark soliton with a domain wall formed at the interface between the spin components of an immiscible binary BEC. The latter can be seen as a solitonic excitation in the sense that it is localized and long lived magnetic excitation in a two-component BEC. The addition of the spin degree of freedom enriches the physics of solitons, both introducing new localized solitonic objects as well as altering the physics of dark solitons.  For example, dark-bright  solitons---a dark soliton in one component whose core is filled with the other component---have been created and collided~\cite{Becker2008}. This is also similar to earlier experiments creating vortices in spinor BECs in which the vortex core was filled with atoms in a different internal state~\cite{PhysRevLett.83.2498}. 

\begin{figure}[]
	\centering
	\includegraphics{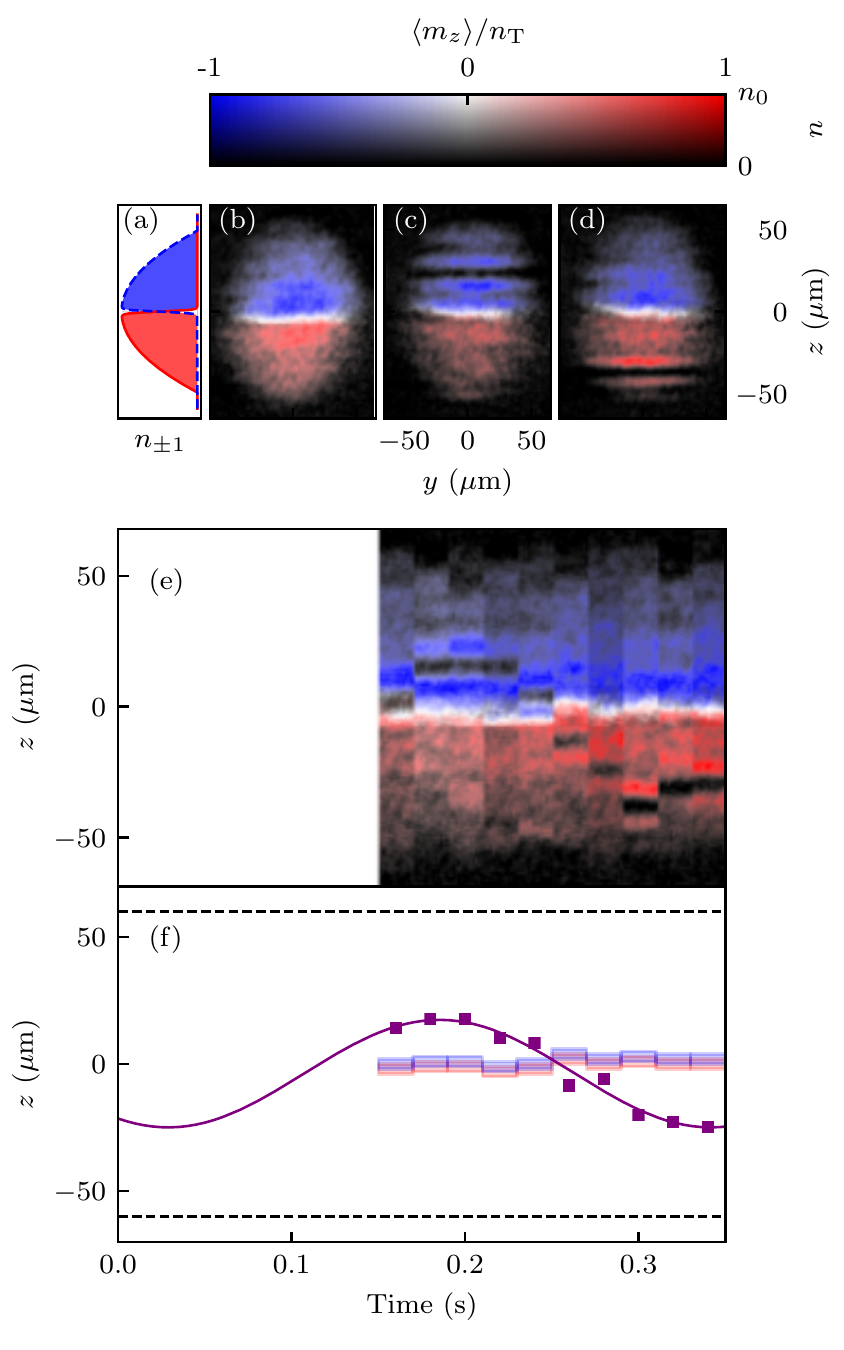}
	\caption{Soliton crossing a domain wall. (a) Domain wall formed by atoms in different spin states. Absorption images of two-component BECs: (b) without a soliton present, (c) with a soliton in the $\vert f = 1, m_f = -1\rangle$ component, and (d) with a soliton in the $\vert f = 1, m_f = +1\rangle$ component. (e) Cross-sectional images resolving the domain wall and the soliton motion. (f) Extracted soliton and domain position averaged over three repetitions of the experiment.}
\label{fig:fig4}
	\end{figure}

To create a domain wall, we apply a radiofrequency $\pi$ pulse to our $\vert f = 1, m_f = 0\rangle$ BEC, putting each atom into an equal superposition of $\vert f = 1, m_f = +1\rangle$ and $\vert f = 1, m_f = -1\rangle$. The small, negative spin-dependent scattering length $a_2$  in $^{87}\text{Rb}$, with $a_2/a \approx -0.005$ makes this binary mixture immiscible~\cite{Ho1998,Widera_2006,Campbell2016}, leading to the formation of stable domain walls with size given by the spin-healing length  $\xi_s = \xi \sqrt{\vert a / a_2\vert} \approx \SI{5}{\mu m}$. The BEC is held for $\SI{2}{s}$ in the presence of a small magnetic field gradient to initialize a domain wall between the two spin states.\footnote{The gradient is not strictly necessary to initiate the domain wall, but the addition of gradient along the longitudinal direction of the BEC decreases the time for the system to reaches equilibrium. We adiabatically apply a bias of the order of $\SI{10^{-4}}{T}$ along the longitudinal direction, which due to imperfection in the alignment is enough to produce the gradient necessary deterministically create the domain wall} After the soliton creation and a chosen propagation time, we perform spin-sensitive imaging by applying a magnetic field gradient ($\approx\SI{0.6}{mT.cm^{-1}}$) during TOF so that atoms in different magnetic states separate before absorption imaging. Figure \ref{fig:fig4}\textcolor{blue}{(a)} we plot the theoretical expected  local density $n_{\pm1}({\bf r})$ of each spin component in a ground-state BEC containing a domain wall, where $n_{+1}({\bf r})$ is displayed in red and $n_{-1}({\bf r})$ is displayed in blue. In Fig. \ref{fig:fig4}\textcolor{blue}{(b)} we plot the magnetization $m_z({\bf r}) \equiv n_{+1}({\bf r}) - n_{-1}({\bf r})$ of such a system after TOF in a color scale normalized by the total density $n_{\rm T}({\bf r}) \equiv n_{+1}({\bf r}) + n_{-1}({\bf r})$.

We create solitons on one side of the domain wall by offsetting the optical potential, patterned by the DMD, from the BEC center. Typical absorption images with solitons on either side of the domain wall are shown in Figs. \ref{fig:fig4}\textcolor{blue}{(c)} and \textcolor{blue}{(d)}. In Fig. \ref{fig:fig4}\textcolor{blue}{(e)} we show a sequence of normalized magnetization slices at different times as the solitons cross the domain wall. Figure \ref{fig:fig4}\textcolor{blue}{(f)} shows the soliton position averaged over three different experimental realizations. From these images we can see that the soliton oscillates as in Fig. \ref{fig:fig2} and is transmitted through the domain wall with no perceptible reflection or change in its trajectory, suggesting that the submicrometer scale soliton travels undistorted through the much thicker domain wall.

\begin{figure}[]
	\centering
	\includegraphics{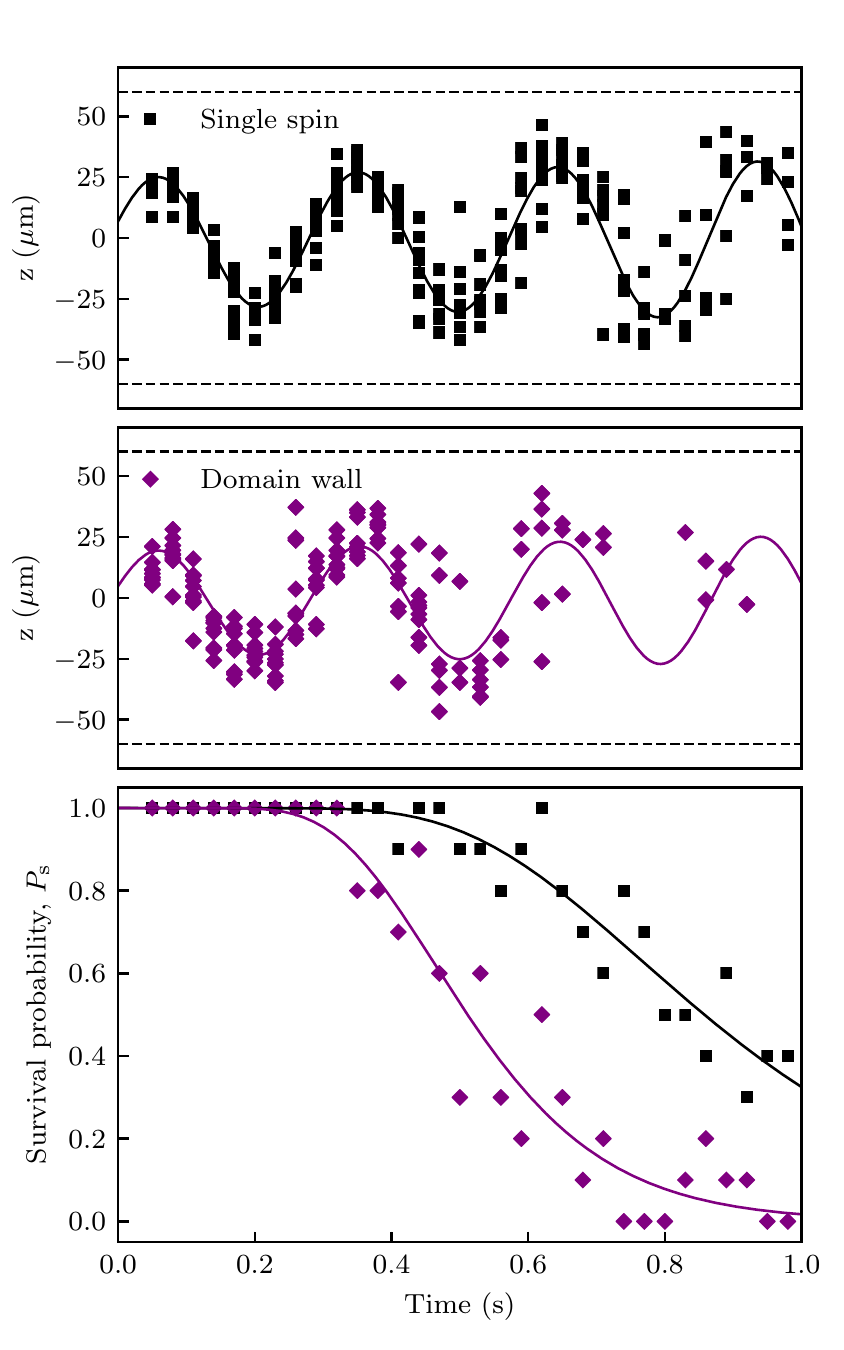}
	\caption{Solitons oscillating in condensates including a domain wall. Position of solitons oscillating in a single-spin condensate (top) and a condensate with a domain wall present (center). For each time there are ten realization of the same experiment, and fewer points indicate that no soliton was observed in one or more realization. Solid curves are fits to $e^{\gamma}\sin(\omega t + \phi)$, where $\gamma$ accounts for the possibility that soliton decay is associated with change in the oscillation amplitude (if these were kink solitons and if the decay mechanism were to involve reduction in the soliton depth, this would increase the soliton velocity and hence its oscillation amplitude). Dashed lines represent the edges of our BECs. The survival probability obtained from these ten realizations is shown in the bottom panel. Solid curves are the fit to the survival probability function.}
\label{fig:fig5}
	\end{figure}

Figure \ref{fig:fig5} shows that the mean soliton trajectory is essentially indistinguishable without (top) and with (middle) the domain wall present, even after multiples passages through the domain wall. Although the mean trajectories are indistinguishable, the fate of solitons differs greatly between these two cases.  The bottom panel shows that successive soliton-domain wall collisions reduce the soliton's survival probability.  In the presence of the domain wall (purple diamonds), the soliton survival probability decays to only $\SI{10}{\%}$ after $\approx\SI{0.7}{s}$, which is much lower than the survival probability for solitons created in single-component BECs (black squares), which is about $\SI{70}{\%}$ for the same time.

Our data shows that solitons are able to cross the domain wall at least two times with a high probability of survival, suggesting that the effects of the passage accumulate, contributing to subsequent decay. We fit (Fig. \ref{fig:fig5} bottom panel) the survival probability to the cumulative lognormal distribution 
\begin{equation}
P_{\rm s}(t) = 1-\frac{1}{2}\erfc\left[-\frac{\ln(t/\tau)}{\sqrt{2}\sigma}\right],
\end{equation}
which describes systems where the probability of decay at time $t$ depends on an accumulation of disrupting events. The constant $\tau$ is the characteristic time when $P_{\rm s}$ falls to $1/2$ and $\sigma$ is the distribution width. In the presence of a domain wall, $\tau_{\rm dw} = \SI{0.51(1)}{s}$ is significantly smaller than for the single-spin case  $\tau = \SI{0.86(2)}{s}$. In both cases the BEC is prepared with the same trap geometry and at a similar temperature so the thermodynamic and transverse instabilities \cite{PhysRevLett.89.110401} should contribute in the same way. One possible dissipation mechanism involves the transfer of energy from the soliton to the domain wall, with each interaction incrementally destabilizing the soliton. Furthermore, the lifetime difference might be explained by a small thermally driven contamination of the majority spin with the minority spin in both domains, as investigated in Refs. \cite{Aycock2503,Hurst2017}. 

Studies with different soliton velocities, soliton sizes, and domain all thicknesses might find a regime where solitons are reflected rather than transmitted when they impact a domain wall.
\section{Outlook and discussion}
We implemented an improved method to create dark solitons in BECs with controlled velocity and arbitrary position, combining standard phase imprinting with density engineering. The observed soliton oscillation frequency along with the theoretical stability diagram for kink solitons~\cite{PhysRevLett.89.110401,Mateo_2015} suggests that we created solitonic vortices; however, our direct experimental evidence is not able to resolve the expected vortex structure~\cite{PhysRevLett.113.065302}. Our numerical simulations in 1D (which is valid with greater transverse frequency or lower chemical potential) show that this technique generates well-controlled kink solitons. 

We demonstrated the utility of the improved method by studying solitons incident on a domain wall and found that solitons pass through the domain wall and have increased decay probability after crossing the domain wall a few times. Even in 1D, collisions between solitons and magnetic domain walls have received little attention, but we expect that they have much in common with soliton-soliton collisions.  When, as in our experiment, the magnetic domain wall is large in comparison with the soliton, we would expect the soliton---an excitation residing in density and phase---to ``adiabatically" follow the slowly changing magnetization. This suggests that the collision will leave the soliton's shape unchanged, but because atoms displaced by the soliton are of opposite spin after the soliton has traversed the domain wall, we expect the domain wall to shift by roughly the healing length in a direction opposite to the soliton propagation direction. This is similar to a ``non interacting'' collision.

When the soliton and domain wall become comparable in size, the soliton can no longer be thought of as a point like particle and a number of outcomes are possible, ranging from non interacting to beam-splitter-like, to perfect reflection. Further studies are needed to fully understand soliton behavior in the presence of a domain wall.

Our capability for creating solitons with tunable velocities, including stationary solitons, enables the study of many phenomena, including dissipative dynamics~\cite{PhysRevLett.116.225301,Hurst2017}, soliton stability and decay in different trap geometries~\cite{PhysRevLett.89.110401,Mateo_2015}, and soliton-soliton collisions in both the ``reflection" and ``transmission" regimes~\cite{PhysRevLett.101.130401,PhysRevLett.101.120406}. Furthermore, high-resolution nondestructive imaging techniques could be used to track the \textit{in-situ} soliton position and better understand their behavior~\cite{Freilich1182}. Lastly, our technique could be used to study the predicted velocity-dependent spin structure of solitons in spin-orbit coupled BECs~\cite{PhysRevA.100.023629}.

\appendix
\section{Calibrating potentials}\label{app:pot}
Our time-of-flight images give a 2D ($y-z$) column density distribution having been integrated along $x$. We separately integrate along $z$ the density distributions on either side of the step, giving us two distributions along $y$. We fit these two distributions to find the two TF radii. We apply the Castin-Dum procedure as if the two sides were separately expanded from our 3D harmonic trap, ignoring the step like potential. We believe this procedure is justified because the confinement, and hence the expansion, is much stronger in the transverse $x,y$ directions than in the longitudinal $z$ direction. From the in-trap TF radii we extract the chemical potential for each side of the step. In the spirit of the local density approximation we identify the difference between these two chemical potentials as the difference of the local chemical potential on either side of the step. According to the Thomas-Fermi approximation, this is the height of the external step potentials.

\section{Numerical simulations}\label{app:gpe}
We simulate the standard and improved phase imprinting methods using a 1D GPE assuming a transverse profile of the inverted-parabola Thomas-Fermi form, with width given by the local chemical potential.  This approach correctly produces the 3D collective mode frequencies when the transverse dynamics are fast, allowing the transverse wavefunction to adiabatically follow the ``chemical potential'' derived from the time-changing 1D density. 

This gives the 1D GPE-like equation
\begin{equation}
i\hbar \frac{\partial}{\partial t}\Psi(z,t) = \left[-\frac{\hbar^2}{2m}\frac{\partial ^2 }{\partial z^2}+U(z)+g'_{\rm 1D}|\Psi(z,t)|\right] \Psi(z,t),
\end{equation}
where $g'_{\rm 1D} = 2\hbar(a\omega_x\omega_y)^{1/2}$, and is related to the usual $g_{\rm 1D} = a^{1/2}g'_{\rm 1D}.$

The time evolution of the GPE is performed through the split-step Fourier method~\cite{Javanainen_2006} in steps of $\SI{0.2}{\mu s}$, and the ground state is found using imaginary time propagation. We simulate ${2.6\times 10^5}$ $^{87}\text{Rb}$ atoms, which for our trap frequencies leads to a chemical potential of $\mu = h\times\SI{1.1}{kHz}$ and $R_{\text{TF}} = \SI{56.4}{\mu m}$. The grid ranges from $\SI{-70}{\mu m}$ to $\SI{70}{\mu m}$ in steps of $\Delta z = \SI{0.025}{\mu m}$.

We account for the finite resolution of the lens system by using potentials with smoothed edges, generated by removing Fourier components above the maximum optical resolution. The patterns used to engineer the density and step potential to imprint the phase are shown in Fig. \ref{fig:fig1}, along with the resolution-limited intensity profile.  We allow the simulated BEC to expand in TOF for $\SI{15}{ms}$ after the soliton oscillation time while quenching the interaction strength to model the rapid Castin-Dum transverse expansion in this 1D simulation. We analyze numerical results in the same way as we analyze the experimental data.

The results for the numerical simulations shown in Fig. \ref{fig:fig3} were obtained using a step potential of magnitude V$_{\rm t} = h\times\SI{5.5}{kHz}$ and a dimple V$_{\rm dt} = h\times\SI{0.78}{kHz}$. Both potentials have the same magnitude used in the experiment. We attribute the small difference between numerical and experimental data to our numerical 1D simulation not fully describing the dynamics in our 3D system.

\section{Distinguishing between slowly moving and stationary solitons}\label{app:fft}
Our experimentally observed soliton positions have uncorrelated shot-to-shot noise on the scale of a few micrometers.  Because most of our solitons oscillate with amplitude much larger than this noise, it generally contributes minimal uncertainty to our fits.  This is not the case for our nearly stationary solitons.

Solitons created with non zero velocity move inside the trap with a characteristic oscillation frequency. In contrast, solitons created at rest do not oscillate but still are susceptible to displacements due to shot-to-shot variation, and a fitting procedure with frequency as a free parameter will misidentify stationary solitons as moving solitons with frequency set by the noise spectrum. To distinguish between these cases we implemented a Fourier spectral analysis. Because our signal undergoes only about three oscillations in our time window of duration $T$,  the $\delta \omega = 2\pi/T$ transform limit makes conventional Fourier methods inapplicable.

For our spectral analysis, we fit the soliton dynamics to $z(t) = A^{\prime}\sin(\omega_{\rm s}^{\prime} t + \phi)$, with the frequency held constant at $\omega_{\rm s}^{\prime}$ for each fit.  The fit-amplitude $A^{\prime}$ provides the spectral weight for that frequency; we then repeat the fitting procedure for different values of $\omega_{\rm s}'$, ranging from $\omega_{\rm s}' = 2\pi\times \SI{1.0}{Hz}$ to $2\pi\times \SI{6.0}{Hz} $ to obtain spectrograms.

\begin{figure}[]
	\centering
	\includegraphics{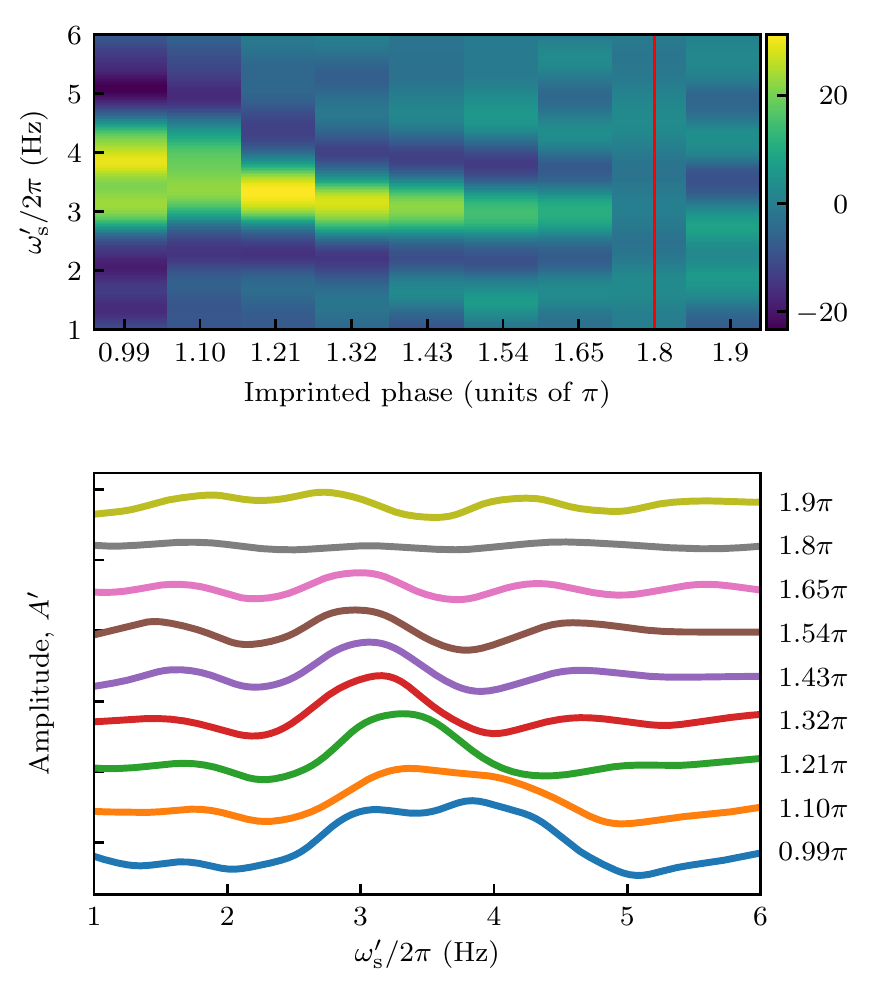}
	\caption{Spectral analysis for solitons created using our improved protocol. Top panel shows in a color scale the amplitude for each frequency $\omega_{\rm s}'$ vs the imprinted phase. Vertical red line is placed at the imprinted phase that creates stationary solitons. Bottom panel shows the individual plots with a vertical displacement to avoid overlapping the curves.}
\label{fig:fft}
	\end{figure}

Figure \ref{fig:fft} shows the results of this technique, applied to solitons produced using our improved method. The top panel shows the amplitude in a color scale, while the bottom panel depicts the data for different applied phases displaced vertically to avoid overlapping the curves.
  
This figure clearly reveals the oscillation frequency at around $2\pi\times \SI{3}{Hz}$ for all but the smallest oscillation, where no motion is resolvable above the noise. For the phase imprint of $1.8\pi$ there is no distinguishable feature at the expected oscillation frequency. From the fit we obtain $A^{\prime} = \SI{0(2)}{\mu m}$ at $\omega_{\rm s}' = 2\pi\times \SI{3}{Hz}$ .

This is the argument underlying our statement that we have created dark solitons with no discernible motion within our experimental uncertainties.  

\begin{acknowledgments}
This work was partially supported by the AFOSR's Quantum Matter MURI, the NIST, and the NSF through the PFC@JQI (Award No. 1430094).  A.M.P. acknowledges support from the Louis Stokes Alliances for Minority Participation (LSAMP) Bridge to the Doctorate Program at the University of Maryland, College Park under Grant No. 1612736.  We are grateful for insights from a very detailed reading of our manuscript by W.~D.~Phillips and M.~Zhao.
\end{acknowledgments}

\end{document}